\documentclass[twocolumn,aps,pra,showpacs]{revtex4-1}
\usepackage{dcolumn}
\usepackage[utf8]{inputenc}
\usepackage{dsfont}
\usepackage{overpic}
\usepackage{caption}    
\usepackage{comment}
\usepackage{graphicx}
\usepackage{amsmath}
\usepackage{upquote,textcomp}
\usepackage{bbold}
\usepackage{subfigure}
\usepackage{caption}
\usepackage{soul}
\usepackage{xcolor}
\usepackage[
  margin=2.5cm,
  includefoot,
  footskip=20pt,
]{geometry}
\usepackage[ colorlinks = true,
             linkcolor = blue,
             urlcolor  = blue,
             citecolor = red,
             anchorcolor = green,
]{hyperref}

\begin{document}
\title{Characterizing quantum detectors by Wigner functions}
\author{Rajveer Nehra\textsuperscript{1}}
\email{rn2hs@virginia.edu}
\author{Kevin Valson Jacob\textsuperscript{2}}
\email{kjaco18@lsu.edu\\}
\affiliation{\textsuperscript{1}Department of Physics, University of Virginia, 382 McCormick Rd, Charlottesville, VA 22904-4714, USA}
\affiliation{\textsuperscript{2}Hearne Institute for Theoretical Physics and Department of Physics \& Astronomy, Louisiana State University,
Baton Rouge, Louisiana 70803, USA }



\begin{abstract}
    We propose a method for characterizing a photodetector by directly reconstructing  the Wigner functions of the detector's Positive-Operator-Value-Measure (POVM) elements. This method extends the works of S. Wallentowitz and Vogel [Phys. Rev. A 53, 4528 (1996)] and Banaszek and Wódkiewicz [Phys. Rev. Lett. 76, 4344 (1996)] for quantum state tomography  via weak-field  homodyne technique to characterize quantum detectors. The scheme uses displaced thermal mixtures as probes to the detector and reconstructs the Wigner function of the photodetector POVM elements from its outcome statistics. In order to make the reconstruction robust to the experimental noise, we use techniques from quadratic convex optimization.
\end{abstract}
\maketitle
\section*{Introduction}
Photodetection has been making consistent progress
 with rapidly developing optical quantum technology~\cite{Hadfield:2009aa,ChristineSilberhorn2007,Wang2017,Ramilli:10, Superconducting_multiplexing_nanowire, TES_2}. Not only do detectors provide us with deeper insights on the quantum behaviour of light by allowing us to perform precise measurements,  they also are an integral part of quantum technology such as quantum computing, quantum enhanced metrology, and quantum communication~\cite{Knill2001, OBrien:2007aa, RevModPhys.79.135, Lloyd1999, clauser_bells_1978,Aspect1982}.

For every quantum detector
, one can associate a set of measurement operators  $\{M_k\}$ called as Positive Operator Value Measures (POVMs). When such a device measures a quantum state $\rho$, the probability of observing an outcome `$k$' is
\begin{equation}\label{eq:povm}
p(k)_\rho=\mathrm{Tr}[\rho M_k].
\end{equation}
Since probabilities are non-negative and sum to one, POVM elements are positive semi-definite and satisfy the completeness property $\sum_{k=0}^{K-1} M_k = \mathbb{1}$. This implies that a set of POVM elements completely describes the measurement device. 
Therefore, in order to characterize a detector, we have to determine its POVM set.

In order to identify the POVM elements of a detector, one can invert Eq.~\eqref{eq:povm} which is known as Quantum Detector Tomography (QDT)~\cite{Lundeen2009NP,ChristineSilberhorn2007}. In optical QDT, light prepared in a set of known tomographically complete states a.k.a probes is incident on the detector to be characterized. The probabilities of different measurement outcomes is then used to characterize the detector.

One such possible set of probes is composed of coherent states $\{|\alpha\rangle \langle \alpha|\}$.
With a coherent state $|\alpha\rangle$ as the probe, the probability of outcome $k$ is given as
 \begin{equation}
p(k)_{|\alpha\rangle} =  \text{Tr}[|\alpha\rangle \langle \alpha | M_k ] =  \pi Q_{M_k}(\alpha ), 
 \label{eq:2}
 \end{equation}
 where $Q_{M_k}(\alpha)$ is the Husimi $Q$ quasi-probability distribution corresponding to the detector POVM element $M_k$. Therefore, one can simply reconstruct the $Q$ functions for POVM elements directly from the measurement statistics. Since $Q_{M_k}(\alpha )$ has complete information about the POVM element $M_k$, ideally it could be used to predict the measurement outcomes for an arbitrary quantum state as follows: consider a quantum state $\rho$  represented in Glauber–Sudarshan $P$ representation as 
 \begin{equation}
 \rho = \int P_\rho(\alpha)  |\alpha\rangle \langle \alpha|  d^2 \alpha,
 \end{equation}
 where $d^2\alpha := d\text{Re}(\alpha)d\text{Im}(\alpha)$. The  probability of outcome $k$ can then be obtained using the Born rule as
 \begin{equation}
p(k)_\rho =  \text{Tr}[\rho M_k ] =  \pi\int P_\rho(\alpha)Q_{M_k}(\alpha )  d^2 \alpha.
 \label{eq:3}
\end{equation}Therefore, by using the $Q$ representation for detector POVM elements and $P$  representation for the input quantum state, one can, in principle determine the outcome probabilities corresponding to detector outcomes.

But this approach suffers from an inherent shortcoming due to the divergent nature of $P$ functions for  nonclassical states of the optical field~\cite{Cahill1969}. In addition, as discussed in~\cite{Lundeen2009NP}, experimental errors and statistical noise during the experiments may distort $Q$ functions resulting in nonphysical POVM elements. In order to  alleviate this shortcoming, it is beneficial to  determine the POVM elements in some basis, for instance photon-number, from the measurement statistics. Several techniques have been proposed and demonstrated to reconstruct the POVM elements in the photon-number basis~\cite{Lundeen2009NP, Grandi_2017, Brida2012, Piacentini:15,Humphreys_2015,Zhang_2012}. One can further represent the POVMs using Wigner quasiprobability distribution functions as discussed in Section~\ref{sec:1}.

Wigner functions, originally introduced by Eugene Wigner in 1932, provide a useful method to visualize quantum states in  phase space~\cite{Wigner,gerry_knight_2004}. The Wigner function corresponding to a quantum state of light has been experimentally obtained using the balanced homodyne method as well as  photon number resolving measurements~\cite{Lvovsky2001, Morin2012, Banaszek1999a, Bondani2010,Sridhar2014a, Nehra2019}.

It is insightful to note that there exists a symmetry between quantum states and measurement operators. We can see this from Eq.~\eqref{eq:povm} wherein, due to the cyclicity of trace,  the roles of the state and the operator can be swapped. This is the underlying relation which we exploit in order to identify the Wigner functions of the measurement operators (POVM elements).

By obtaining the Wigner functions of the detector, any experimental probability can be found in terms of the Wigner functions of the state as well as of the detector, which are well-behaved unlike $P$ functions. Thus, Eq.~\eqref{eq:3} can be written as
\begin{equation}
    p(k)_\rho = \text{Tr}[\rho M_k] = \int W_\rho(\alpha)W_{M_k}(\alpha )  d^2 \alpha,
\end{equation}
where $W_\rho(\alpha)$ and $W_{M_k}(\alpha )$ are the Wigner functions of quantum state $\rho$ and POVM element $M_k$ respectively.

In this work, we propose an alternative method for QDT by directly reconstructing the Wigner quasiprobability functions corresponding to detector POVM elements: it alleviates the need of finding the POVMs in the photon-number basis.  Apart from the fundamental interest in obtaining Wigner functions of a detector, our scheme is particularly beneficial to study the decoherence of a quantum detector by observing the behaviour of the POVM Wigner functions in certain regions of the phase space~\cite{PhysRevLett.107.050504}. 

This paper is organized as follows. In section~\ref{sec:1}, we detail the method for characterizing photodetectors using displaced thermal mixtures. In section ~\ref{sec:2}, we then apply this method to photon-number-resolving detectors. Section ~\ref{sec:3} discusses the  resources  required for characterizing a phase-insensitive detector. We discuss the number of phase space points where Wigner function should be experimentally measured in order to have a good confidence in reconstruction. In section~\ref{sec:4}, we use convex optimization techniques to make our reconstruction robust to noise.  Finally, we note our conclusions in section~\ref{sec:5}.
\section{Method}\label{sec:1}
We use the well known result that the Wigner function operator can be represented in Fock space as
\begin{equation}
\hat{W}(\alpha)=\frac{2}{\pi}\sum_{n=0}^\infty (-1)^n \hat{D}(\alpha)|n\rangle\langle n|\hat{D}^\dagger(\alpha), 
\end{equation}
where $\hat{D}(\alpha) = \exp({ \alpha\hat{a}^\dagger - \alpha^*\hat{a}})$ is the displacement operator with $\alpha \in \mathbb{C}$~\cite{Royer1977}. 
For a detector, our aim is to characterize the Wigner functions corresponding to its POVM elements. Since POVMs are self-adjoint positive semi-definite operators, one can write the Wigner function of a POVM element $M_k$ as
\begin{equation}
    W_{M_k}(\alpha)=\frac{2}{\pi}\sum_{n=0}^\infty (-1)^n \mathrm{Tr}\left[M_k\hat{D}(\alpha)|n\rangle\langle n|\hat{D}^\dagger(\alpha)\right],\label{eq:sum}
\end{equation} 
where, for simplicity, we define 
\begin{equation}
    P_{M_k}^{(n)}(\alpha) := \mathrm{Tr}\left[M_k\hat{D}(\alpha)|n\rangle\langle n|\hat{D}^\dagger(\alpha)\right].
    \label{Eq:P}
\end{equation}
Although the sum in Eq.~\eqref{eq:sum} has infinite terms but in practice one can truncate it to `$n_0$' as  further terms don't significantly contribute to the sum. Thus, we have 
\begin{align}
W_{M_k}(\alpha) 
&\approx \frac{2}{\pi}\sum_{n = 0 }^{n_0}(-1)^n P_{M_k}^{(n)}(\alpha).
\label{eq:sum1}
\end{align} 
From Eq.~\eqref{eq:sum1} we can see that finding the Wigner function corresponding to `$M_k$' amounts to finding out all these summands. 

In this paper, we will restrict ourselves to phase-insensitive detectors for simplicity. Such detectors have the Wigner functions of their POVM elements rotationally symmetric around the origin, and hence can be characterized on the real line alone. However, we note that this scheme is applicable to phase-sensitive detectors also; and for such detectors, we have to choose $\alpha$ in the complex plane.
\begin{figure}[!hbt]
    \includegraphics[width=0.5\textwidth]{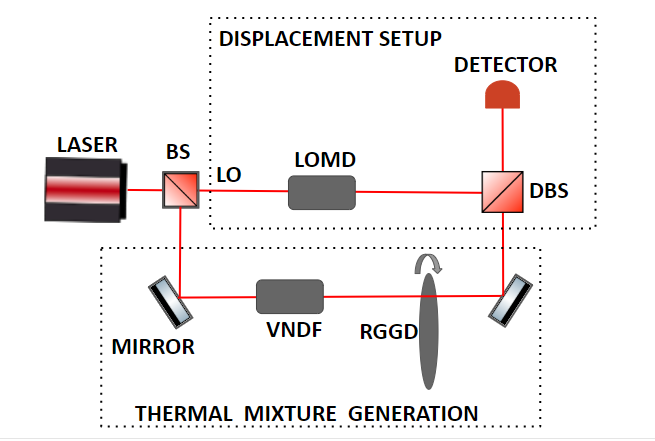}
    \caption{Schematic of the experimental setup.}
    \label{fig:setup}
\end{figure}
\subsection{Proposed experimental setup}
 
Fig.~\ref{fig:setup} shows a schematic for our proposed experiment. A laser beam is split into two beams at the first beamsplitter. One beam is used to generate thermal mixtures. Thermal mixtures can be generated by randomzing the phase and amplitude of the laser beam (coherent state). To achieve that, we use a Variable Neutral Density Filter (VNDF) along with a Rotating Ground-Glass Disk (RGGD). These allow us to generate thermal mixtures with a variable mean photon-number~\cite{Thermalstate}. The other beam is used as a Local Oscillator (LO) for phase space displacements. For displacements, we interfere thermal mixtures with the LO at a highly unbalanced beamsplitter denoted as Displacement beamsplitter (DBS) in the experiment schematic. A local oscillator modulator (LOMD) is used for varying the displacements, which are required to probe of the Wigner function over the whole phase space.

In order to do this, we consider $(n_0 +1)$ distinct thermal mixtures given as
\begin{equation}
    \rho^{(j)}=\sum_{n=0}^\infty p_n^{(j)}|n\rangle\langle n|
\end{equation}
where $j$ = $0,\hdots,n_0$ labels the thermal states, and $p_n^{(j)} = \frac{\Bar{n}^n_j}{(1+\Bar{n}_j)^{n+1}}$ is the Bose-Einstein photon-number distribution of a thermal mixture $\rho^{(j)}$ with mean photon-number $\Bar{n}_j$. We then displace these thermal mixtures by amplitude $\alpha$ which is, in general, a complex number. Then, the probability of obtaining `$k$' outcome with the displaced thermal input as input is give by
\begin{equation}
    Q_k^{(j)}(\alpha) \approx \sum_{n=0}^{n_0} p_n^{(j)} P_{M_k}^{(n)}(\alpha),
    \label{eq:Q}
\end{equation}
where we choose the thermal state such that the contribution to the RHS from the omitted terms is negligible. This is possible because the thermal state has an exponentially decreasing photon number distribution. In matrix form, we can write Eq.~\eqref{eq:Q} as

\begin{equation}
\begin{pmatrix}
Q_k^{(0)}(\alpha) \\
Q_k^{(1)}(\alpha) \\
\vdots\\
Q_k^{(n_0)}(\alpha)
\end{pmatrix}=\begin{pmatrix}
p_0^{(0)} & p_1^{(0)} & \hdots & p_{n_0}^{(0)} \\
p_0^{(1)} & p_1^{(1)} & \hdots & p_{n_0}^{(1)} \\
\vdots\\
p_0^{(n_0)} & p_1^{(n_0)} & \hdots & p_{n_0}^{(n_0)}
\end{pmatrix}
\begin{pmatrix}
 P_{M_k}^{(0)}(\alpha) \\
 P_{M_k}^{(1)}(\alpha) \\
 \vdots \\
 P_{M_k}^{(n_0)}(\alpha)
\end{pmatrix}.
\label{eq:matrix}
\end{equation}
We can further write Eq.~\eqref{eq:matrix} compactly as
\begin{equation}
  \bf{Q =  P\Pi^\alpha _{M_k}},
  \label{eq:Qcompact}
\end{equation}
where $\bf{Q}$ and $\bf{\Pi^\alpha_{M_k}}$ are vectors of length $(n_0+1)$, and $\bf{P}$ is the probability distribution square matrix of dimension $(n_0+1)\times(n_0+1)$. 
Thus by solving Eq.~\eqref{eq:Qcompact}, we can determine $\bf{\Pi^\alpha_{M_k}}$, which allows us to calculate the summation in Eq.~\eqref {eq:sum}.
To solve for $\bf{\Pi^\alpha_{M_k}}$, one needs to  solve the following convex quadratic optimization problem:

\begin{align}
\text{Minimize}\enspace&\{\bf{||Q - P\Pi^\alpha _{M_k}||_2}\}, \nonumber\\
\begin{split}
\text{Subject to}&\enspace\bigg\{ 1{\bf{\geq\Pi^\alpha _{M_k}}}\geq 0, \enspace\\
&-1\leq\sum_{n=0}^{n_0}(-1)^n P^{(n)}_{M_k}(\alpha)\leq1\bigg\},
\label{eq:const}
\end{split}
\end{align} 
where $||.||$ is the $l_2$ norm defined as  $||V||_2 = \big(\sum_{i}|V_{i}|^2\big)^{1/2}$ for a vector $V$.

The optimization constraints in Eq.~\eqref{eq:const} can be understood as follows. First, the $n^{th}$ element $P^{(n)}_{M_k}(\alpha)$ of {$\bf{\Pi^\alpha _{M_k}}$} is essentially the probability of getting $k$-click if a displaced $n$-photon Fock state is incident to the detector. Therefore, we have $1\geq P^{(n)}_{M_k}(\alpha)\geq 0$. 
Second, Wigner functions are well- defined and bounded between $[-2/\pi, 2/\pi]$  for a POVM element corresponding to a phase-insensitive detector. This is because the POVM element in such a case is a statistical mixture of projectors. In this case we can use the second constraint in Eq.~\eqref{eq:const}.

Thus solving this optimization allows us to determine the Wigner function at a given phase space point $\alpha$. Further, we can repeat the process with different displacement amplitudes to reconstruct the Wigner function over the entire phase space. In practice, the Wigner functions of various detectors are localized around the origin and vanishes to zero for large $\alpha$, so it is unnecessary to displace the thermal states with arbitrarily large $\alpha$. 

In the following section we numerically simulate this method for a phase-insensitive detector. We hasten to add that this method is applicable to any type of detector. 

\section{Modelling a Photon-number-resolving detector}\label{sec:2}
\begin{figure*}[hbt!]
\centerline{\includegraphics[width=\textwidth]{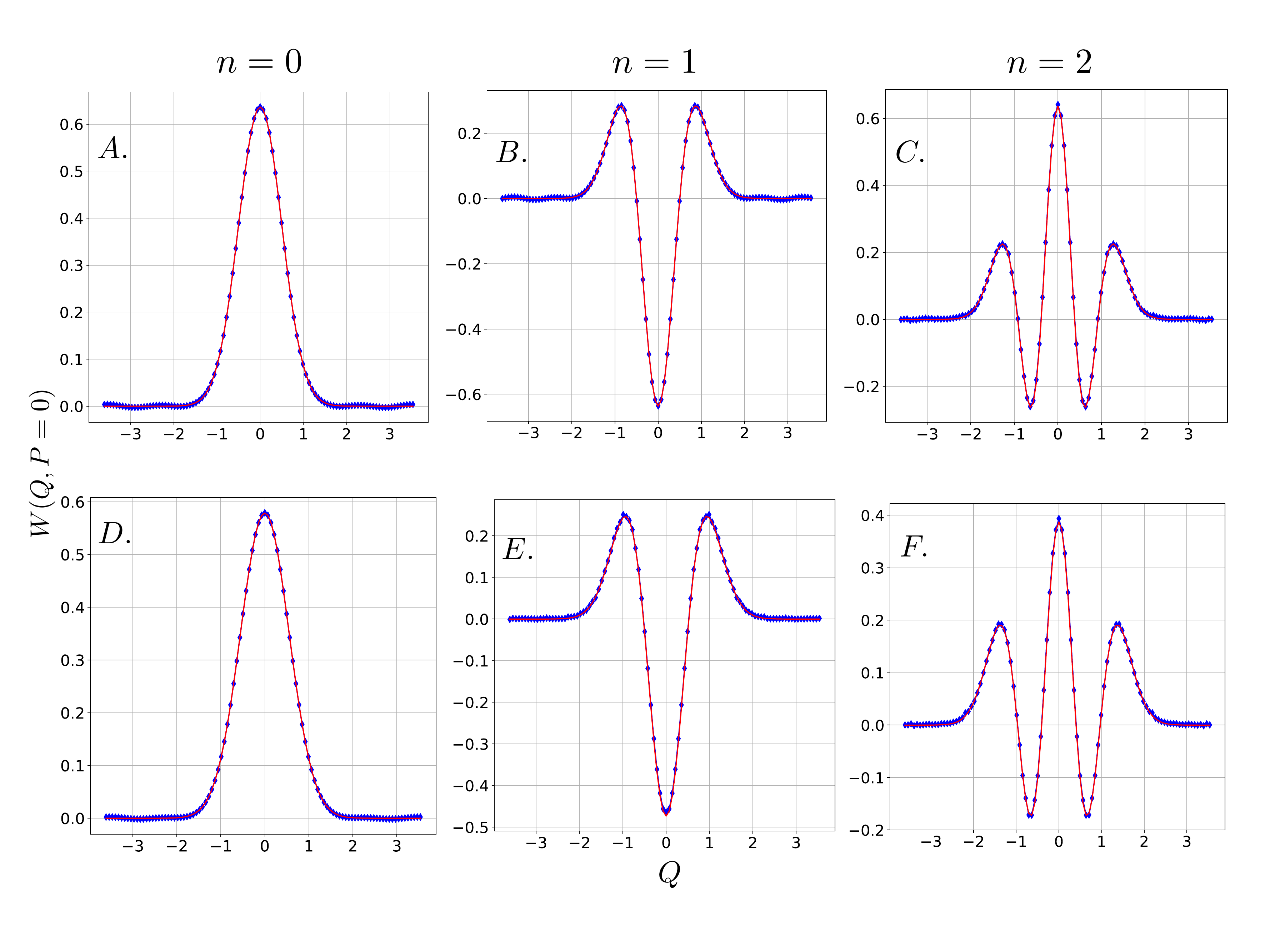}}
\caption{Wigner functions for POVM elements corresponding to zero-, one- and two-photon detection events. Red curves are theoretically expected Wigner functions and blue ones the reconstructed one using the proposed method. In the top row, A, B, and C are for a perfect PNR detector; and in the bottom row D, E, and F are for a PNR detector with  detection efficiency $\eta = 0.90$}. 
    \label{fig:povm_pnr}
\end{figure*}
In this section, we reconstruct the Wigner functions of a perfect and an imperfect photon-number-resolving (PNR) detector. In general, a POVM element corresponding to `$k$' outcome can be written in the photon-number basis as
\begin{equation}
M_k = \sum_{m,n=0}^{\infty}\langle m|M_k|n\rangle|m\rangle\langle n|, 
   \label{eq:genPOVM}
\end{equation} 
where $\langle m|M_k|n\rangle$ are the matrix elements of the POVM operator. One can further simplify Eq.~\eqref{eq:genPOVM} for a PNR detector with no dark counts as
\begin{align}\label{eq:povm_pnr}
   M_k  = \sum_{m=k}^{m_0}\langle m|M_k|m\rangle|m\rangle\langle m|.
\end{align}

Note that Eq.~\eqref{eq:povm_pnr} differs from Eq.~\eqref{eq:genPOVM} in three ways. First, the POVM is diagonal with entries $\langle m|M_k|m\rangle$, which are essentially the probabilities of detecting `$k$' photons given `$m$' photons are incident to the detector. Thus for a detector with detection efficiency $\eta$, we have
\begin{equation}
    p(k|m) = \langle m|M_k|m\rangle= {m\choose k}\eta^k(1-\eta)^{m-k}.
    \label{eq:prob_pnr}
\end{equation}
Second, we have truncated the sum to `$m_0$' such that it exceeds the photon-number at which saturates the detector. Third, the sum is starting from `$k$' because with no dark counts noise, one would expect `$k$' clicks only if there are $m\geq k$ photons are incident on the detector.  

Eq.~\eqref{eq:povm_pnr} and Eq.~\eqref{eq:prob_pnr} can be interpreted as follows: The POVM elements of a perfect PNR detector are  projectors $\Pi _m = |m\rangle \langle m|$. However, for an imperfect detector, its  efficiency $\eta < 1$.  If $m$ photons impinge on such a detector, due to its non-unity detection efficiency, $k<m$ photons results in a detection event contributing a factor of $\eta^k$ to the probability of the event; while $(m-k)$ photons remain undetected contributing a factor of $(1-\eta)^{m-k}$ to the probability of the event. Thus, such POVMs are statistical mixtures of projective measurements.

In numerical simulations, we considered equidistant 51 displacement amplitudes in $\alpha \in [-3.6, 3.6]$, which allowed us to probe the Wigner function uniformly over the entire region of phase space where the Wigner function is non-vanishing. We used 50 equally spaced thermal states of mean photon-number in $\bar{n}\in[0,4]$. For all of our simulations in open source Python module QuTip~\cite{Johansson2013QuTiP2A}, we generally limited the dimension of the Hilbert space to 50, and the sum in Eq.~\eqref{eq:povm_pnr} was truncated with $m_0 = 50$ at which point $P(k|m_0)$ was of the order of $10^{-10}$ for $\eta = 0.90$.

In Fig.~\ref{fig:povm_pnr}, we plot the Wigner functions of one, two, and three photon detections for a perfect detector and an imperfect detector with imperfections as modelled in Eq.~\eqref{eq:povm_pnr}. From Fig.~\ref{fig:povm_pnr}, we notice that the extrema of the Wigner functions of imperfect detectors are closer to the origin than those of perfect detectors. This is due to the contribution of higher order projectors in the Wigner functions of imperfect detectors. In particular for the imperfect single-photon detection event, we see a reduced negativity in the Wigner function around the origin. This is due to the contribution of the Wigner function of the two-photon detection event which is strongly positive around origin. Similar arguments can be made for the the reduced positivity of the Wigner function for zero- and two-photon detection event POVMs.


In our reconstruction, we have uniformly sampled the phase space. A natural question that now arises is whether the number of points that needs to be probed in the phase space can be reduced. We investigate this question in the following section. 

\begin{figure*}%
    \centering
  {{\includegraphics[width=0.48\textwidth]{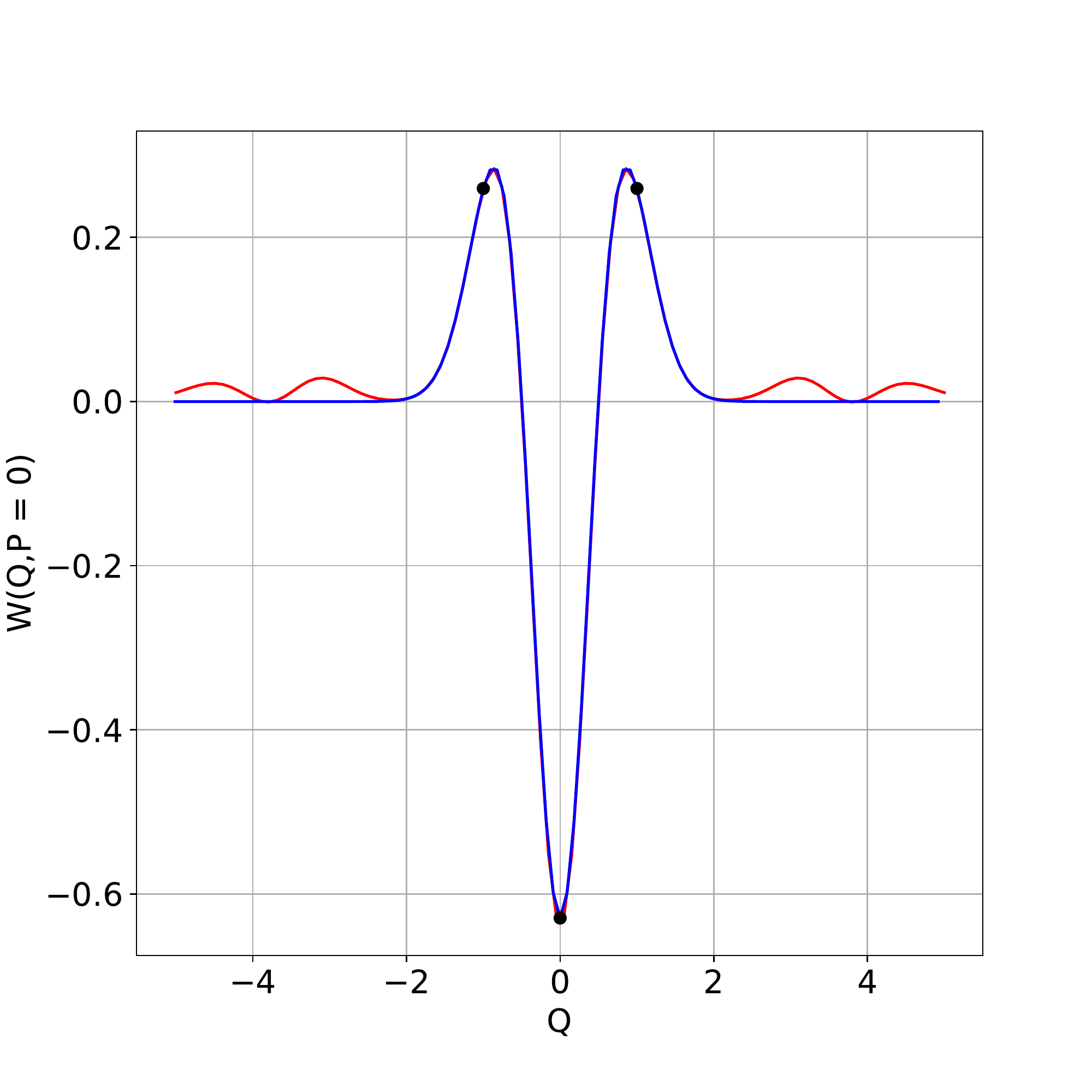} }}%
{{\includegraphics[width=0.48\textwidth]{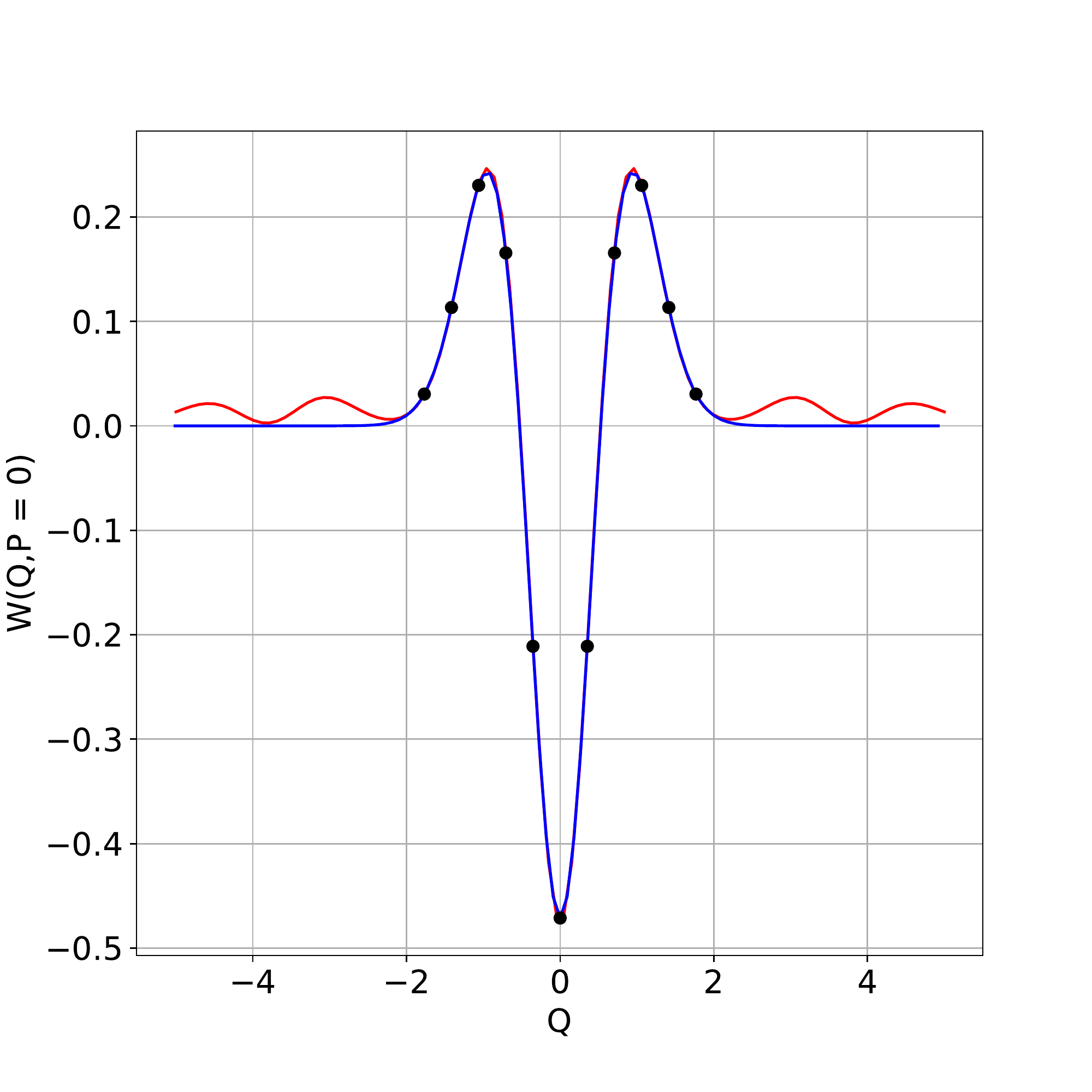} }}%
\caption{Left : Wigner function corresponding to a perfect single-photon detection POVM determined by naive summation up to 15 terms of Eq.~\eqref{eq:sum} (Red), and using an Gaussian modulated quadratic fit near the origin (Blue). Black points represent the phase space points where the Wigner function was probed by the proposed method here. The latter approximates well the actual Wigner function. Right : Wigner function corresponding to an imperfect single-photon detection POVM with $\eta = 0.90$. }
    \label{fig:polynomial_reconstruction}
\end{figure*}
\section{characterizing phase-insensitive detectors with polynomial resources}\label{sec:3}
Although the method outline earlier is general, it had substantial resource requirements as we had to uniformly sample over the phase space. However, this requirement can be drastically reduced if we have the prior knowledge that the detector is phase-insensitive, i.e. the representations of its POVM elements are diagonal in the Fock basis. Note that the phase sensitivity of a detector can easily be checked by varying the phase of the LO while keeping the amplitude fixed. In this case, unlike a PNR detector, a phase sensitive detector outputs different measurement statistics for different phases and fixed amplitudes of the LO.

We recall that the Wigner functions of Fock states are Gaussian modulated Laguerre polynomials~\cite{Leonhardt1997}. This allows us to write the Wigner function the POVM elememt `$M_k$' of a PNR detector as
\begin{equation}
    W_{M_k}(\alpha) = \frac{2e^{-2|\alpha|^2}}{\pi}\sum_{m=0}^{m_0} (-1)^m p(k|m) \  {\mathcal{L}}_m(4|\alpha|^2),
   \label{eq:insensitive}
\end{equation}
where ${\mathcal{L}}_m(x)$ represents the Laguerre polynomial of $m^{th}$ degree in $|\alpha|^2$. As the Wigner function is a function of $|\alpha|^2$, it is symmetric around the origin, and can be fully characterized on the real line. Since the Wigner function is a Gaussian modulated polynomial, the problem of reconstructing it is reduced to finding out a polynomial of degree $2m_0$ in $\alpha$ which requires us to find the Wigner function only at $2m_0 +1$ points. 

As an example, we considered the POVM element corresponding to a single-photon detection event for both perfect and imperfect PNR detectors. In Fig.~\ref{fig:polynomial_reconstruction}, the red curves show the POVM determined by the naive summation up to 15 terms of Eq.~\eqref{eq:sum}; and the blue curves the reconstructed Wigner functions with black points being the phase space coordinates where the Wigner function was probed.

We see that one needs to probe the Wigner function only at three points for a perfect detector because the Laguerre polynomial ${\mathcal{L}}_{m=1}(4|\alpha|^2)$ is quadratic in $\alpha$, and therefore can be fully characterized using three distinct points. Likewise, the Wigner function for an imperfect single-photon POVM can be reconstructed using only 11 distinct points (black points in Fig.~\ref{fig:polynomial_reconstruction}) if we truncate the sum in Eq.~\eqref{eq:povm_pnr} at $m_0 = 5$ where $p(k|m)$ is of the order of $10^{-6}$. In this case, we will have to reconstruct an Gaussian modulated polynomials of degree 10 because the last term in the Eq.~\eqref{eq:povm_pnr} would be a projector, $|5\rangle \langle 5|$ with Wigner function given by Gaussian modulation of ${\mathcal{L}}_{m=5}(4|\alpha|^2)$. 

Note that finding the Gaussian modulated polynomial also works for a general detector given by Eq.~\eqref{eq:genPOVM}. However, instead of reconstructing the Wigner function on the real line, we will have to reconstruct \textcolor{blue}{it} in the complex plane for which appropriate polynomial interpolation schemes have to be used \cite{Gasca2000}.

\section{Robustness against experimental noise}\label{sec:4} 

In this section, we discuss the robustness of this method against experimental noise. In general, inverting Eq.~\eqref{eq:Qcompact} is ill-conditioned as seen by the large ratio of the largest and smallest singular values of the matrix $\bf{P}$. This makes the reconstructed POVM elements extremely sensitive to small fluctuations in the measurement statistics, and can lead to nonphysical POVMs. 

However, the effects of ill-conditioning can be remarkably suppressed by adding a regularization to the optimization problem. Several types of regularization techniques are discussed in detail in \cite{Lundeen2009NP}, and for this work we use Tikhonov regularization~\cite{Tik63}. Using this technique, inverting Eq.~\eqref{eq:Qcompact} can be mathematically formulated as the following optimization problem:


\begin{align}
\text{Minimize}\enspace&\{{||\bf{Q - P\Pi^\alpha _{M_k}}||_2 +\gamma||\bf{\Pi^\alpha _{M_k}||_2}}\}, \nonumber\\
\begin{split}
\text{Subject to}&\enspace\bigg\{ 1\geq{\bf{\Pi^\alpha _{M_k}}}\geq 0, \enspace\\   &-1\leq\sum_{n=0}^{n_0}(-1)^n P^{(n)}_{M_k}(\alpha)\leq1\bigg\},
\end{split}
\end{align}
where $\gamma$ is the regularization parameter. Solving this problem translates to a convex quadratic optimization which can be efficiently solved using a semi-definite problem solver, for instance, the Python package CVXOPT. 
\begin{figure*}[!hbt]
    \centering
    \includegraphics[width =0.9\textwidth]{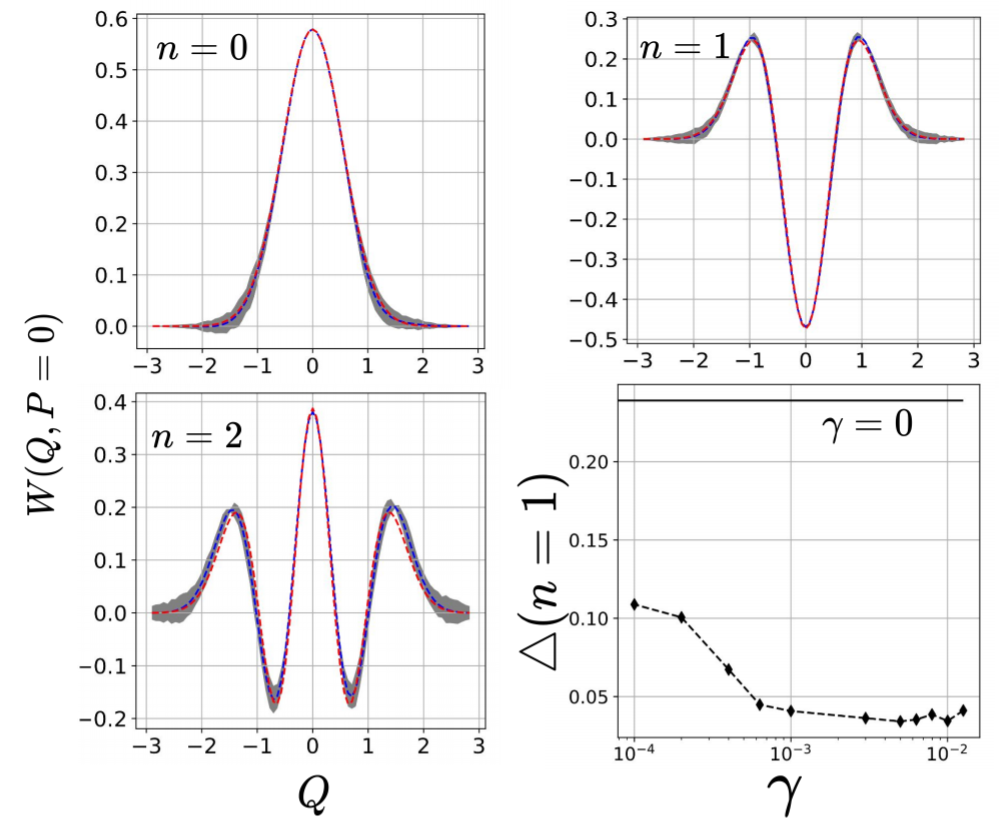}
    \caption{Blue: Reconstructed Wigner functions using regularization for zero-,one- and two-photon detection event of a detector with $\eta = 1$. Red curves are theoretically expected Wigner functions. Gray areas are error $(1\sigma)$ obtained using $N = 40$ iterations. Bottom right: Dashed-diamond curve illustrates the robustness of the reconstruction against regularization parameter $\gamma$ and black solid line is without regularization, i.e, $\gamma =0$.}
    \label{fig:robustpovm}
\end{figure*}
In order to simulate the presence of noise in our reconstruction, we introduce noise in the LO's amplitude $|\alpha|$. We model this noise as a Gaussian distribution of mean zero and standard deviation $\sigma = 1\%|\alpha|^2$. This is the typical noise level present in currently available stabilized lasers.  Therefore, the displacement amplitudes are $(\alpha_1+\delta d_1, \alpha_2+\delta d_2  \dots , \alpha_{max} +\delta d_{max})$, where each $\delta d_i$ is a random variable sampled from the Gaussian distribution. To further reduce the effects of the fluctuations, we average the Wigner functions obtained over $N = 40$ iterations of the optimization. As a result, we get 
\begin{equation}
    \overline{W}_{{M_k}}(\alpha) = \frac{\sum_{j = 1}^{N} W^{j}_{{M_k}}(\alpha+\delta{\alpha}_j)}{N}. 
\end{equation}
Having obtained $\overline{W}_{{M_k}}(\alpha)$, we then we utilize robust nonlinear regression methods to further suppress the fluctuations. We recall that for a phase insensitive detector, the POVMs are Gaussian modulated polynomials of degree $2m_0$ in $\alpha$, where $m_0$ is the saturation limit given in Eq.~\eqref{eq:insensitive}. Therefore, once we have experimentally probed the Wigner function at $2m_0 + 1 $ distinct points of the phase space, we could simply fit a Gaussian modulated polynomial of degree $2m_0$ in $\alpha$ to reconstruct the Wigner function over the entire phase space. Keeping that in mind, we set an optimization problem as: 
\begin{align*}
  &\text{Minimize:}\\ &\left\{\frac{1}{2}\sum_{i=1}^{}L\left[\left(e^{-2|\alpha_i|^2}\text{Poly}(2m_0,\alpha_i) - \overline{W}_{{M_k}}(\alpha_i)\right)^2\right]\right\}, 
\end{align*} 
where $L$ is defined as 
\begin{equation}
    L(y) = 2(\sqrt{1+y} - 1),  
\end{equation}
and $\text{Poly}(2m_0,\alpha_i)$ is a polynomial of degree $2m_0$. Note that this approach of finding the Gaussian modulated polynomial has an advantage of not being biased unlike the simple least-square fitting method which tends to significantly bias in order to avoid high residuals in the data \cite{Robust_convex}. 

We further evaluate the quality of reconstruction method by using the relative error defined with $l_2$ norm as:
\begin{equation}
    \triangle : = \frac{||W^{theory}_{M_k}(\alpha) - W^{reconstruted}_{M_k}(\alpha)||_2}{||W^{theory}_{M_k}(\alpha)||_2}
    \label{eq:FOM}
\end{equation}

The result of our reconstruction is shown in Fig.~\ref{fig:robustpovm}. Since the fluctuations grow with increasing local oscillator amplitude, the reconstruction of the Wigner function around the origin of phase space is the least disturbed, but with higher displacements the fluctuations grow stronger as seen in Fig.~\ref{fig:robustpovm}. Therefore, it may be beneficial to probe the Wigner function around the origin densely, and sparsely at the higher displacements, in particular $|\alpha|>1$. Note that probing near the origin doesn't undermine the quality of reconstruction as long as we probe the Wigner function at $2m_0+1$ distinct points because we need only $2m_0 + 1$ distinct points to reconstruct a polynomial of degree $2m_0$ as seen in Fig.~\ref{fig:polynomial_reconstruction}. 

In fact, we can further exploit the rotational symmetry of the POVMs corresponding to phase insensitive detector, which means the Wigner function at $\alpha$ has the same value at $-\alpha$. This allows us to only probe the Wigner function at $m_0+1$ distinct points to fully characterize a quantum detector that saturates at the photon-number $m_0$. However, in this work we numerically probe the phase space at equidistant displacement amplitudes.

We now investigate how sensitive our reconstruction is to the choice of $\gamma$. To evaluate that, we calculate the relative error defined in Eq.~\eqref{eq:FOM} for several values of $\gamma \in$ $[10^{-4}, 0.012]$. The result is illustrated on the bottom right in Fig.~\ref{fig:robustpovm} for the POVM element corresponding to $n=1$ and $\eta = 0.90$. We can clearly see that even if we vary $\gamma$ by an order of magnitude (from $10^{-3}$ to $10^{-2}$), the relative error only changes by less than one percent. This shows that there is sufficient freedom in the choice of $\gamma$.\\

\section{Conclusions}\label{sec:5}
We have developed a method for characterizing photodetectors by experimentally obtaining the Wigner functions corresponding to the POVMs describing the detector measurements. The proposed experimental scheme is simple and easily accessible, in particular, for phase insensitive detector. Augmented with convex quadratic optimization and robust nonlinear fitting techniques, we demonstrated its robustness to the experimental fluctuations. 

Future work on this method may involve an account for mode mismatch between the local oscillator and the optical mode of thermal mixtures. This direction of research is motivated by the fact that unlike in the balanced homodyne technique, mode mismatch cannot simply be treated as losses in this method.

\section*{Acknowledgements}
KVJ acknowledges the support from Louisiana State University System Board of Regents via an Economic
Development Assistantship. RN was supported by NSF grant PHY-1708023. The authors are grateful to Prof. Olivier Pfister and Carlos Andres Gonzalez Arciniegas for the feedback on the manuscript. RN is indebted to Prof. Jonathan P. Dowling for financially supporting the visit to LSU.
The authors thank Miller Eaton, Austin P. Lund, Arshdeep Sekhon, and Vikesh Siddhu for helpful discussions.\\

The authors contributed equally to this work.





\bibliographystyle{apsrev4-1}
\bibliography{PNR_ref,Pfister,quantumoptics}
\end{document}